\documentclass[traditabstract]{aa} 
\usepackage{graphicx}
\usepackage{amsmath}

\usepackage{txfonts}
\begin{document}
   \title{Disentangling between stellar activity and planetary signals}

   \author{I. Boisse
          \inst{1,2}
          \and
          F. Bouchy\inst{1,3}\fnmsep
          \and
          G. H\'ebrard\inst{1,3}
          \and
          X. Bonfils\inst{4,5}     
          \and 
           N. Santos\inst{2}  
           \and 
          S. Vauclair\inst{6} 
            }

   \institute{Institut d'Astrophysique de Paris, Universit\'e Pierre et Marie Curie, UMR7095 CNRS, 98bis bd. Arago, 75014 Paris, France\\
              \email{iboisse@iap.fr}
            \and
            Centro de Astrof{\'\i}sica, Universidade do Porto, Rua das Estrelas, 4150-762 Porto, Portugal
             \and
              Observatoire de Haute Provence, CNRS/OAMP, 04870 St Michel l'Observatoire, France
         \and
           Laboratoire d'Astrophysique de Grenoble, Observatoire de Grenoble, Universit\'e Joseph Fourier, CNRS, UMR 5571, 38041, Grenoble Cedex 09, France 
           \and
           Observatoire de Gen\`eve, Universit\'e de Gen\`eve, 51 Ch. des Maillettes, 1290 Sauverny, 
Switzerland       
               \and 
        LATT-UMR 5572, CNRS \& Universit\'e P. Sabatier, 14 Av. E. Belin, F-31400 Toulouse, France
  }

   \date{Received ; accepted }

 
  \abstract
   { Photospheric stellar activity (i.e. dark spots or bright plages) might be an important source of noise and confusion in stellar radial-velocity (RV) measurements. Radial-velocimetry planet search surveys as well as follow-up of photometric transit surveys require a deeper understanding and characterization of the effects of stellar activities to differentiate them from planetary signals.  
   We simulate dark spots on a rotating stellar photosphere. The variations in the photometry, RV, and spectral line shapes are characterized and analyzed according to the stellar inclination, the latitude, and the number of spots. 
   We show that the anti-correlation between RV and bisector span, known to be a signature of activity, requires a good sampling to be resolved when there are several spots on the photosphere. The Lomb-Scargle periodograms of the RV variations induced by activity present power at the rotational period $P_{rot}$ of the star and its two first harmonics $P_{rot}/2$ and $P_{rot}/3$. Three adjusted sinusoids fixed at the fundamental period and its two-first harmonics allow us to remove about 90\% of the RV jitter amplitude. We apply and validate our approach on four known active planet-host stars: HD\,189733, GJ\,674, CoRoT-7, and $\iota$\,Hor. We succeed in fitting simultaneously activity and planetary signals on GJ674 and CoRoT-7. This simultaneous modeling of the activity and planetary parameters leads to slightly higher masses of CoRoT-7b and c of respectively, 5.7 $\pm$ 2.5 M$_{Earth}$ and 13.1 $\pm$ 4.1 M$_{Earth}$. The larger uncertainties properly take into account the stellar active jitter. We exclude short-period low-mass exoplanets around $\iota$\,Hor. For data with realistic time-sampling and white Gaussian noise, we use simulations to show that our approach is effective in distinguishing reflex-motion due to a planetary companion and stellar-activity-induced RV variations provided that 1) the planetary orbital period is not close to that of the stellar rotation or one of its two first harmonics, 2) the semi-amplitude of the planet exceeds $\sim$30\% of the semi-amplitude of the active signal, 3) the rotational period of the star is accurately known, and 4) the data cover more than one stellar rotational period.
   }


   \maketitle

\section{Introduction}

	High-precision radial-velocimetry has until now been the most efficient way of discovering planetary systems. However, an active star displays on its photosphere dark spots and/or bright plages that rotate with the star. These inhomogeneities across the stellar surface can induce RV shifts due to changes in the spectral line shape that may add confusions with the Doppler reflex-motion due to a planetary companion (e.g. Queloz et al. 2001; Desidera et al. 2004; Hu\'elamo et al. 2008). The amplitude of the RV shifts depend on the $v$\,$\sin$\,$i$ of the star, the spectrograph resolution and both the size and temperature of spot (Saar \& Donahue, 1997; Hatzes 1999; Desort et al. 2007). Stellar activity generates RV jitter and prevents accurate measurements of the stellar motions. For these reasons, active stars are then usually discarded from RV surveys using criteria based on the activity index $R^{'}_{HK}$ and/or $v$\,$\sin$\,$i$. 
However, photometric transit-search missions (such as CoRoT and Kepler) require RV measurements to establish the planetary nature of the transiting candidates and 
 characterize their true masses. These candidates include active stars adding strong confusions and difficulties in the RV follow-up.\\

The anti-correlation that exists between RV and bisector span (the measurement of the spectral line asymmetry) is currently used as a signature of RV variations induced by stellar activity (e.g. Queloz et al. 2001; Hatzes et al. 2002). Removing this anti-correlation to subtract the activity-induced RV may allow us to derive more accurate parameters of the planetary system (e.g. Melo et al. 2007; Boisse et al. 2009). Another approach to distinguishing stellar activity and planetary signatures consists of adjusting two Keplerian signals, one with the planetary period and the second with the stellar rotational period (e.g. Bonfils et al. 2007; Forveille et al. 2008).\\

In these past few years, the impact of the stellar activity has been studied. 
Desort et al. (2007) quantified how the impact of stellar spots on RV measurements depended on the spectral type, the rotational velocity, and the characteristics of the spectrograph used. Santos et al. (2009) investigated long-term HARPS measurements of a sample of stars with known activity cycles. Lagrange et al. (2010) and Meunier et al. (2010) analyzed spots, plages, and convection simulations computed using solar data.\\ 

In this paper, we characterize variations in all the observed 
parameters derived from RV measurements as well as from photometric measurements due 
to stellar activity. We study how the Keplerian fit used to search for planets 
in RV data is affected by spots and we test an approach to subtract RV jitter based on harmonic decompositions of the star rotation. For this, we use simulations of spectroscopic measurements of rotating spotted stars and validate our approach on active stars monitored by high-precision spectrographs: HD\,189733 (Boisse et al. 2009), GJ\,674 (Bonfils et al. 2007), CoRoT-7 (Queloz et al. 2009), and $\iota$\,Hor (Vauclair et al. 2008).

\section{Simulations of activity-induced radial velocity}
\label{simulations}

\subsection{SOAP tool: dark spot simulations}

SOAP (Boisse et al., in prep.) is a program that calculates the photometric, RV, and line shape modulations induced by one (or more) cool spots on a rotating stellar surface. SOAP computes the rotational broadening of a spectral line by sampling the stellar disk on a grid. For each grid cell, a Gaussian function represents the typical line of the emergent spectrum. The Gaussian is Doppler-shifted according to the projected rotational velocity ($v$\,$\sin$\,$i$) and weighted by a linear limb-darkening law. The stellar spectrum output by the program is the sum of all contributions from all grid cells. The spot is considered as a dark surface without any emission of light, so we cannot compute different temperatures for the spot. For a given spot (defined by its latitude, longitude, and size), SOAP computes which grid cells are obscured and removes their contribution to the integrated stellar spectrum. 
	
   \begin{figure}[b]
   \centering
   \includegraphics[width=9cm]{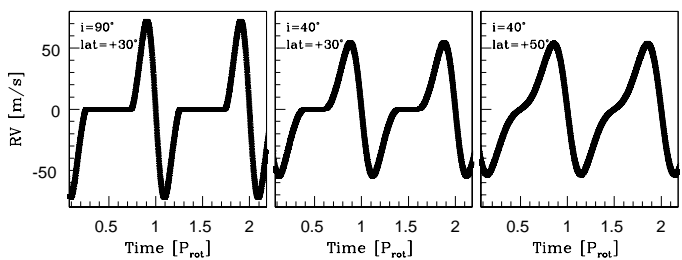}
      \caption{ RV modulations due to one spot as a function of time (expressed in rotational period unit). At $t$=0, the dark spot of 1\% of the visible stellar surface is in front of the line of sight. The shape of the signal changes with the inclination $i$ of the star and the latitude $lat$ of the spot, labelled in the top left of each panel. 
              }
         \label{myfig1}
   \end{figure}

For the simulation, we choose a G0V star with a radius of 1.1 $R_{\sun}$ and a $v$\,$\sin$\,$i$=5.7\ kms$^{-1}$, a linear coefficient of the limb darkening of 0.6, and a spectrograph resolution of 
110 000 in order to be in the same conditions of the $\iota$ Hor data presented in Sect. 3.4. We fixed arbitrarily a dark spot size of 1\% of the visible stellar surface. The stellar spectrum output by SOAP is an averaged spectral line, equivalent to the cross-correlation function (CCF) computed to measure RV with real data. We fit the simulated CCF with a Gaussian that sets the parameters of the CCF (RV, contrast, FWHM, and bisector span). The photometric flux is also computed.

\subsection{RV variations due to a dark spot}
\label{RVspot}	
Fig.~\ref{myfig1} shows the RV modulations due to a spot as a function of time for different inclinations $i$ of the star with the line of sight and different spot 
latitudes $lat$. These two parameters clearly modify the pattern of the RV modulation. 
If the spot remains visible during all the stellar rotation ($lat \geqslant i$), the shape is close to a sinusoidal function  (Fig.~\ref{myfig1}, \textit{right}). If the spot is hidden during the rotation of the star (Fig.~\ref{myfig1}, \textit{left}), the RV variation resembles a Rossiter-McLaughlin (RM) effect (Rossiter 1924; McLaughlin 1924). 
	
\begin{figure}[h]
   \centering
   \includegraphics[width=8.5cm]{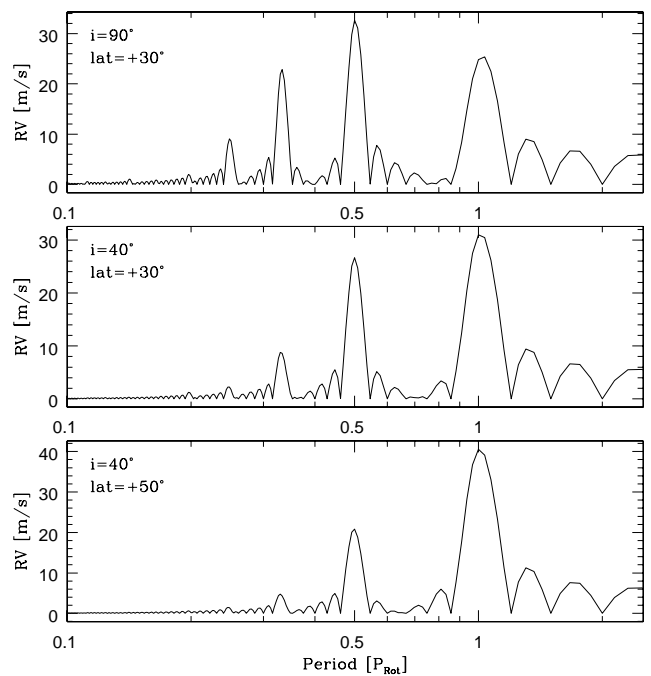}
      \caption{Lomb-Scargle periodograms of the three RV modulations showed in Fig.~\ref{myfig1}. The fundamental frequency, $P_{rot}$, and its first harmonics are detected. 
         }
         \label{myfig2}
   \end{figure}

Fig.~\ref{myfig2} shows the Lomb-Scargle periodograms of the three cases showed in Fig.~\ref{myfig1}. Main peaks are clearly detected at the rotational period of the star $P_{rot}$, as well as the two first harmonics $P_{rot}/2$ and $P_{rot}/3$. We note that the energy in each peak varies with the shape of the RV modulation. Multiples of the rotational period are never found. Low-amplitude signal is detected at $P_{rot}/4$ but only when the star is seen equator-on and the spot is close to the equator. In that case, the RV change departs strongly from a sinusoidal shape and the periodogram exhibits a stronger amplitude excess at $P_{rot}/2$ rather than at the stellar rotational period. 

\begin{figure}[h]
   \centering
   \includegraphics[width=8.7cm]{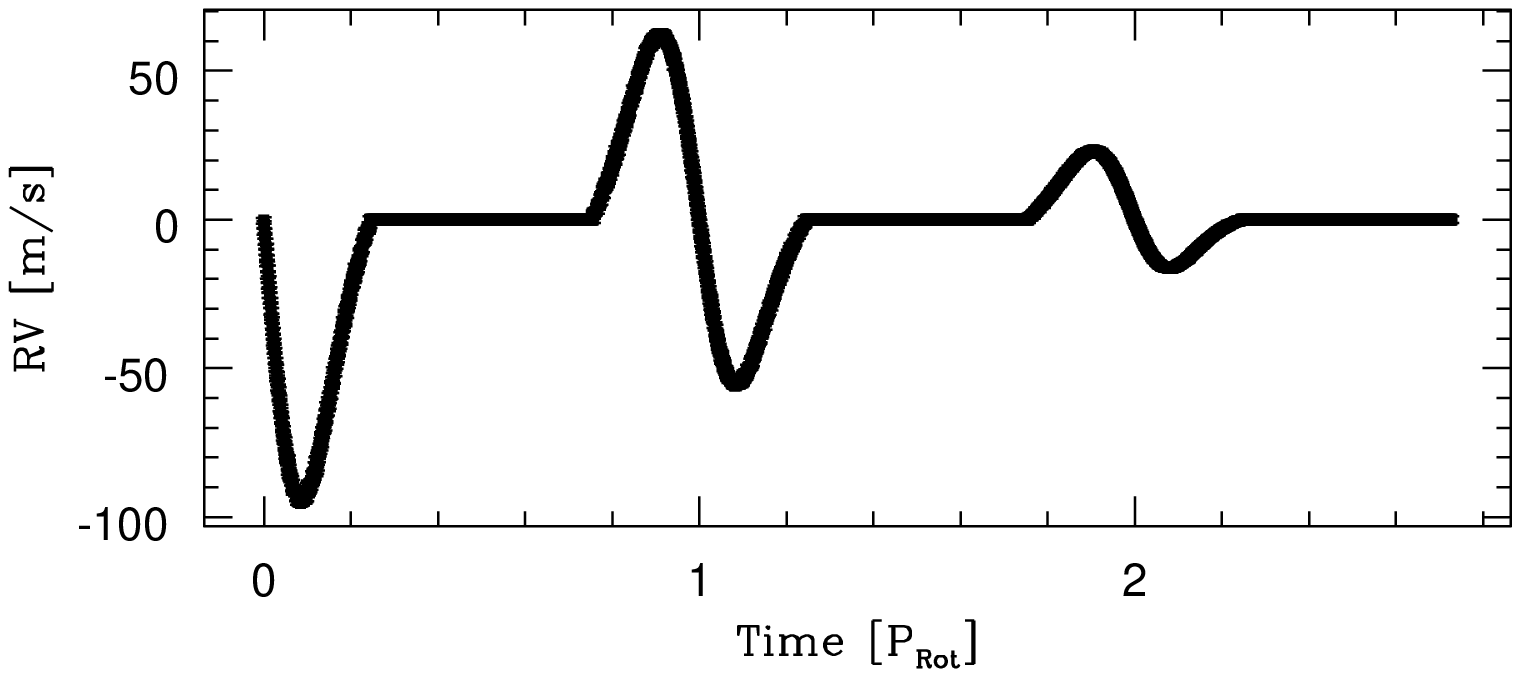}
    \includegraphics[width=8.4cm]{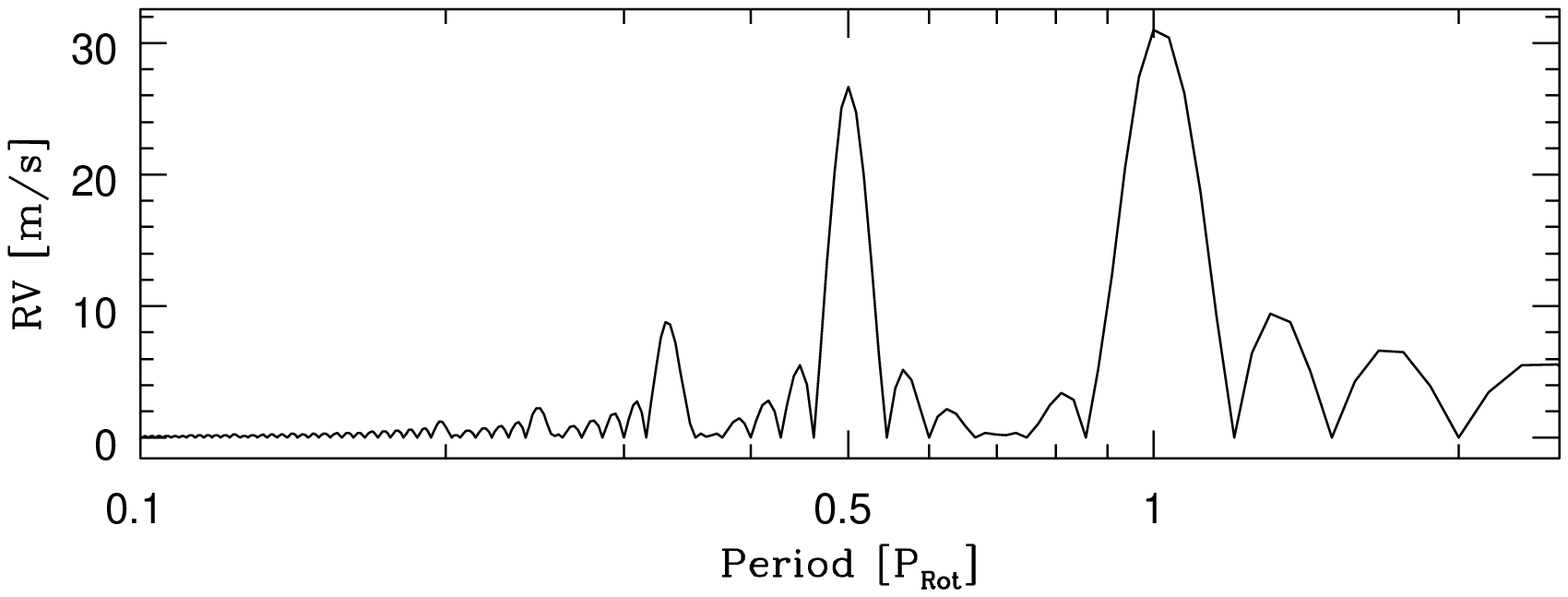}
         \caption{\textit{Top:} RV as a function of time when the spot size decreases. \textit{Bottom:} Lomb-Scargle periodogram of the RV. Peaks correspond to the rotational period and its harmonics.}
         \label{myfig3}
   \end{figure}

A third of the active regions, where spots grow and decay, appear at the same location as a previous active region and their lifetime can be several rotation timescales (e.g. Howard 1996). Hence, the phase of the RV jitter is preserved when the spot movement is only linked to the stellar rotation. In Fig.~\ref{myfig3}, we simulated the RV modulation taking into account the evolution of a spot, i.e. when the spot size and/or temperature changes with time. The Lomb-Scargle periodogram has identical peaks at $P_{rot}$ and its two first harmonics. Finally, the periods detected in the periodogram are the same for the following configurations: 1) a star with different inclinations, 2) spots at different latitudes, 3) spot size varying with time, and 4) several spots on the stellar surface (cf.~Sect.~\ref{severalspots}).

\subsection{CCF parameter variation due to a dark spot}
\label{ccfparameters}

The bisector span (BIS) is a measurement of the asymmetry of the CCF, which corresponds more or less to the average line of the spectrum. An anti-correlation between the RV and the BIS is a signature of activity-induced RV variations. The slope of the anti-correlation depends on the spot size, the $v$\,$\sin$\,$i$ of the star and the resolution of the instrument (Santos et al. 2003; Desort et al. 2007). 

\begin{figure}
   \centering
   \includegraphics[width=8.cm]{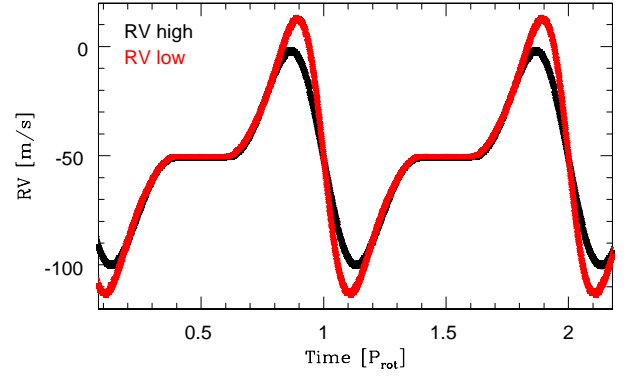}
      \caption{ RV$_{high}$ (black) and RV$_{low}$ (red) as a function of time. The star has an inclination $i$=40$^{\circ}$ and the latitude of the spot is equal to +30$^{\circ}$. RV$_{low}$ is more sensitive to activity than RV$_{high}$.
              }
         \label{myfig9}
   \end{figure}

\begin{figure}
   \centering
   \includegraphics[width=9.cm]{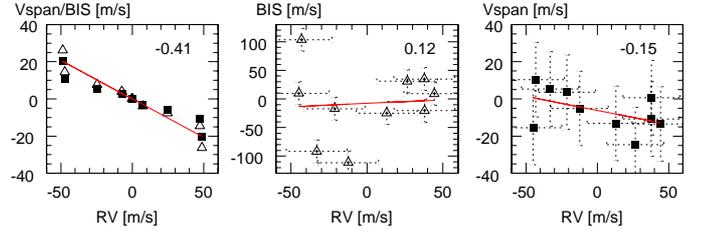}
      \caption{ \textit{Left:} BIS (triangles) and V$_{span}$ (squares) as a function of RV for a simulated spot rotating at latitude +50$^{\circ}$ on a star with an inclination $i$=40$^{\circ}$. \textit{Middle and Right:} Respectively, BIS (Queloz et al. 2001) and V$_{span}$ (this paper) as a function of RV for the same simulated spot with 20ms$^{-1}$ additional photonic noise in the CCF. The lines are the least squares fit to the data. The numbers in the right-hand corner is the value of the slope of the fit.
              }
         \label{myfig11}
   \end{figure}

The BIS was defined by Queloz et al. (2001) by computing the RV difference of the upper and the lower part of the bisector. This measurement appears unreliable at low signal-to-noise ratio (SNR). We describe here another approach to estimate the asymmetry of a CCF. 
We fit the CCF in two steps selecting the upper and bottom part to compute 
the RV$_{high}$ and RV$_{low}$, respectively. Where $\sigma$ is the width of the CCF, the upper part of the CCF is defined in the range [-$\infty$:-1$\sigma$][+1$\sigma$:+$\infty$] and the lower part is defined 
in the range given by [-$\infty$:-3$\sigma$][-1$\sigma$:+1$\sigma$][+3$\sigma$:+$\infty$]. The RV$_{low}$ is more sensitive to the variation in the bottom of the line, and then, more sensitive to stellar activity as shown in Fig.~\ref{myfig9}. The difference RV$_{high}$\,$-$\,RV$_{low}$\,=\,V$_{span}$ gives a measurement of the CCF asymmetry. In Fig.~\ref{myfig11}, the two techniques of CCF asymmetry measurement are plotted for two regimes of SNR illustrating that at low SNR, the V$_{span}$ is more robust than BIS.  
	 
\begin{figure}
   \centering
   \includegraphics[width=9.cm]{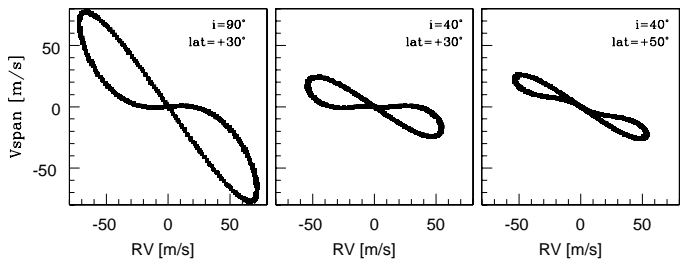}
      \caption{ V$_{span}$ as a function of RV for the three RV modulations of Fig.1. The relation is not a perfect anti-correlation but looks more like an inclined {\it eight} shape.
              }
         \label{myfig11b}
   \end{figure}

For one spot, the V$_{span}$ or BIS is anti-correlated with the RV as expected (see Fig.~\ref{myfig11b}). Although the shape of the anti-correlation is not a straight line, but a loop more or less like an inclined {\it eight}. Depending on the geometrical configuration, the mean 
slope and the width of the $eight$-like loop vary.

The subtraction of the average slope of the anti-correlation has previously been used to remove the activity-induced RV in order to improve the parameters of a Keplerian planetary orbit (e.g. Melo et al. 2007; Boisse et al. 2009). We simulated the effect of this approach and found that the pattern of the RV residuals after the correction is identical with a lower amplitude of the 
initial RVs. The detected periods in the Lomb-Scargle periodogram of the RV residuals 
are indeed the same, with an amplitude ratio equivalent to those in the initial RVs periodogram. 
The use of the RV-BIS anti-correlation to correct RV jitter has several limitations.
First, it merely reduces the amplitude of the jitter signal, and does not alter its
frequency content. Second, it can only be used if the $v sin i$ of the star is
sufficient to resolve the effect of spots on the bisector (Desort et al. 2007;
Bonfils et al. 2007). Third, we have shown that the degree of correlation depends on
the spot latitude and stellar inclination. Finally, the correlation may be absent if
 effects other than activity contribute significantly to the RV variations (Queloz et
al. 2009).

\begin{figure}
   \centering
   \includegraphics[width=8.4cm]{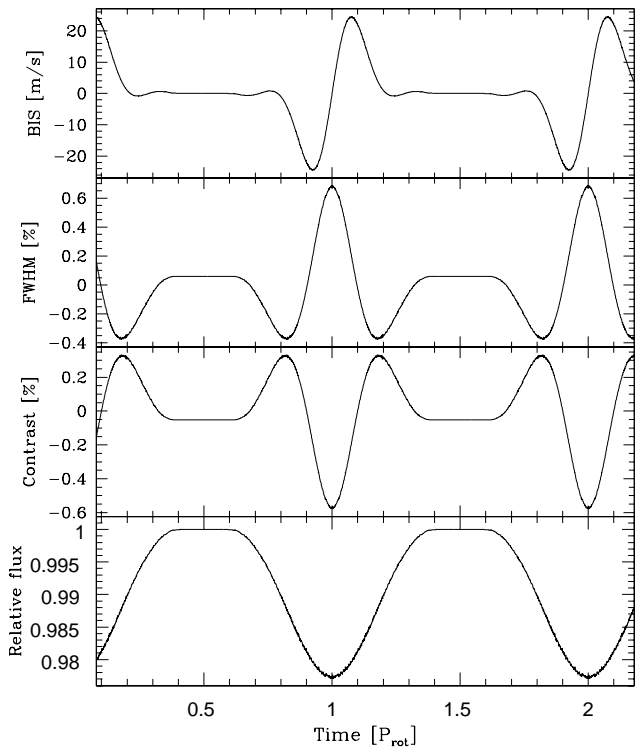}
      \caption{\textit{From top to bottom:} \textit{BIS}, FWHM, contrast of the CFF and the photometric flux as a function of time. The variation in the parameters of the CCF are expressed as a percentage of the variation around the mean value. The star has an inclination $i$=40$^{\circ}$ and the latitude of the spot is equal to +30$^{\circ}$ (RV shown in Fig.~\ref{myfig1} \textit{middle}). At $t$=1, the spot is in front of the line of sight.
              }
         \label{myfig10}
   \end{figure}

Our simulation does not take into account the variation in line width with stellar effective temperature, which is affected by the presence of spots as measured and calibrated by Santos et al. (2002). In our simulated CCF, the flux contribution of the area occulted by the spot is subtracted from the mean stellar line. The resulted CCF is distorted relative to a Gaussian profile. As the spot rotates with the stellar surface, the contribution of the spot moves inside the CCF, from the blue to the red wings. The contrast and FWHM of the fitted CCF vary in anti-correlation as illustrated in Fig.~\ref{myfig10} where the CCF parameters are expressed as a percentage of variation relative to the mean value. When the spot is in front of the line of sight, its contribution affects the center of the line profile, the contrast is reduced and the FWHM is larger. In contrast, when the spot contribution is in the wings of the line, the FWHM is smaller. In practice, CCF contrast may be more sensitive to both the conditions of observation (e.g. sky background) and the diffuse light in the spectrograph. The CCF FWHM is then probably the most reliable parameter to characterize an effect due to the activity. 
	

\subsection{Correlation with photometry variations}	

\begin{figure}
   \centering
   \includegraphics[width=9cm]{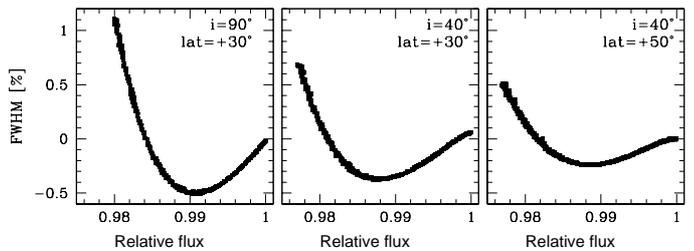}
      \caption{ FWHM of the CCF as a function of the photometric flux for the three cases of Fig.~\ref{myfig1}. The variations in the FWHM are expressed as a percentage of the variation around the mean value (here, equal to zero).
              }
         \label{myfig12}
   \end{figure}

We compared the variations in the photometric flux with those of the CCF parameters (asymmetry, width, and contrast) for different stellar inclinations and spot latitudes (see Fig.~\ref{myfig10}). The contrast is anti-correlated with the FWHM. Two regimes are observed 
between the CCF contrast and the photometric flux (see Fig.~\ref{myfig12}).
 The simulations compute the distortion of the CCF as a lack of flux masked by the spot. When the spot is in front of the line of sight, i.e. when the emitted flux is at its lowest, there is a bump at the line center and the wings appear broader. In contrast, when the spot is in the limb, i.e. when the distortion is in the wings and the emitted flux is at its medium value, the line width appears to be narrower. This explains why an anti-correlation is observed between CCF width and photometry. The other part, where a correlation is observed between the parameters is not well understood. As shown in Fig.~\ref{myfig10}, the contrast and FWHM vary in anti-correlation, the effect in terms of FWHM then also depends on the contrast variations. This part of the relation between FWHM and photometry might originate from our inability to compute the different temperatures of the spot. 
On Corot-7 (Queloz et al. 2009), the photometric measurements from the Swiss Euler telescope are anti-correlated with the HARPS CCF width.

\subsection{RV fit of a dark spot}
\label{rvfitofadarkspot}

We now attempt to remove or at least reduce the stellar activity signals in order to identify a planetary signal hidden in the RV jitter. We add to the simulated RV time series a Gaussian noise of $\sigma$=1 ms$^{-1}$ to take into account photon noise and/or instrumental noise. The main shape of the RV modulations and the peaks in the periodograms are not affected with such a noise having an amplitude 100 times lower than the activity jitter. 

\begin{figure}[h]
   \centering
   \includegraphics[width=8.4cm]{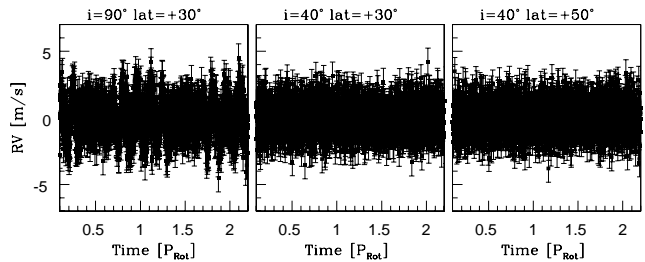}
        \caption{Residuals from the five-sinusoid fit as a function of time of the RV variations showed in Fig.~\ref{myfig1}. 
              }
         \label{myfig7}
   \end{figure}

\begin{figure}[h]
   \centering
    \includegraphics[width=8.4cm]{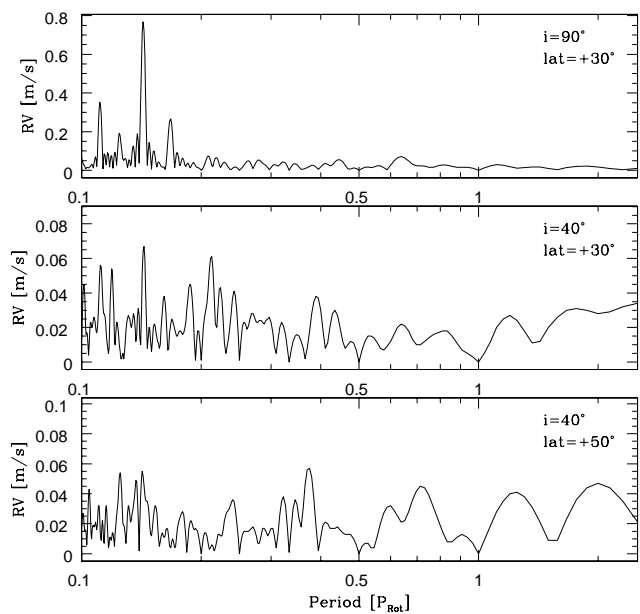}
      \caption{ Lomb-Scargle periodogram of the residuals of the 3 RV variations showed in Fig.~\ref{myfig1} after the five-sinusoid fit with fixed periods at the rotational period of the star and its harmonics.  
              }
         \label{myfig6}
   \end{figure}

The Lomb-Scargle periodogram corresponds to sinusoidal decompositions of the data. We saw previously that the main signal are at the stellar rotational period $P_{rot}$ and its two first harmonics $P_{rot}/2$ and $P_{rot}/3$. In some configurations, signal is present at $P_{rot}/4$. We firstly fit five sinusoids to the three simulated cases of Fig.~\ref{myfig1} with periods fixed at the rotational period $P_{rot}$, and its first harmonics. The residuals are plotted in Fig.~\ref{myfig7}, and the Lomb-Scargle periodogram of the residuals in Fig.~\ref{myfig6}. The parameters of the sinusoids fitted are reported in Table~1. The active jitter is reduced from more than 100 ms$^{-1}$ peak-to-peak to less than 3 ms$^{-1}$. In reality, only three sinusoids are needed to decrease the semi-amplitude of the RV jitter by more than 87\% because the semi-amplitude of the fourth ($P_{rot}/4$) and fifth ($P_{rot}/5$) sinusoids are negligible in almost all cases.

We compared the fit of activity signals with Keplerian and sinusoids. Fitting a Keplerian is equivalent to fitting a sinusoid at the same period and its first harmonics. The parameters of the fit are reported in Table~2. The number of free parameters is equal to 16 for a fit with three Keplerians and to 10 with three sinusoids. Taking into account only the fundamental period $P_{rot}$ as a free parameter (and not the other periods fixed as harmonics), there are 14 free parameters for a fit with three Keplerians against 8 for a fit with three sinusoids. Moreover, if
some energy due to activity remains in the residuals, it is at the next harmonic for a fit with sinusoids, while it may be at any other harmonic for a fit with Keplerians. This is why we choose to fit the active jitter with three sinusoids fixed at the stellar rotational period and its two first harmonics.

 The latest harmonics may appear because the sampling of the simulation is almost perfect. In the following section, we checked that in practice when the sampling and coverage of the stellar activity signal are limited, a three sinusoid fit with fixed periods at the rotation period, and its two first harmonics is sufficient.\\ 	

\begin{table*}[h]
  \centering 
  \caption{Sinusoidal orbital solutions for fitting RV jitter activity for different inclinations of the star and latitudes of one stellar spot (Fig.~\ref{myfig1}).}
  \label{param_p}
\begin{tabular}{ccccccc}
\hline
\hline
& Parameters&&& \\
\hline
$i$=90$^{\circ}$ and lat=+30$^{\circ}$ & $P_{P}$   & $P_{rot}$ & $P_{rot}/2$ & $P_{rot}/3$ & $P_{rot}/4$ & $P_{rot}/5$\\
& K  [\,m\,s$^{-1}$]     &  24.83 $\pm$ 0.02 & 32.65 $\pm$ 0.02 & 22.78 $\pm$ 0.02  & 8.66 $\pm$ 0.02 & 1.14 $\pm$ 0.03\\
& $\sigma_{(O-C)}$ [\,m\,s$^{-1}$]  &    &before the fit : 33.77 & after the fit : 1.26; $\chi^{2}$ = 1.22\\
\hline
\\
$i$=40$^{\circ}$ and lat=+30$^{\circ}$ & $P_{P}$   &  $P_{rot}$ & $P_{rot}/2$ & $P_{rot}/3$ & $P_{rot}/4$ & $P_{rot}/5$ \\
& K  [\,m\,s$^{-1}$]     &  30.97 $\pm$ 0.02 & 26.68 $\pm$ 0.02 & 8.57 $\pm$ 0.02 & 1.79 $\pm$ 0.04 & 0.85 $\pm$ 0.03\\
& $\sigma_{(O-C)}$ [\,m\,s$^{-1}$]  &     & before the fit : 29.58 & after the fit : 1.05; $\chi^{2}$ = 1.01\\
\hline
\\
$i$=40$^{\circ}$ and lat=+50$^{\circ}$ & $P_{P}$  &      $P_{rot}$ & $P_{rot}/2$ & $P_{rot}/3$ & $P_{rot}/4$ & $P_{rot}/5$ \\
& K  [\,m\,s$^{-1}$]     &  40.59 $\pm$ 0.02 & 20.80 $\pm$ 0.02 & 4.29 $\pm$ 0.02 & 0.97 $\pm$ 0.02 & 0.18 $\pm$ 0.02\\
& $\sigma_{(O-C)}$ [\,m\,s$^{-1}$]  &     & before the fit : 32.41 & after the fit : 1.04; $\chi^{2}$ = 1.00\\
\hline
\end{tabular}
\end{table*}

\subsection{Several dark spots}
\label{severalspots}
\begin{figure} 
\begin{center}
\includegraphics[width=8.4cm]{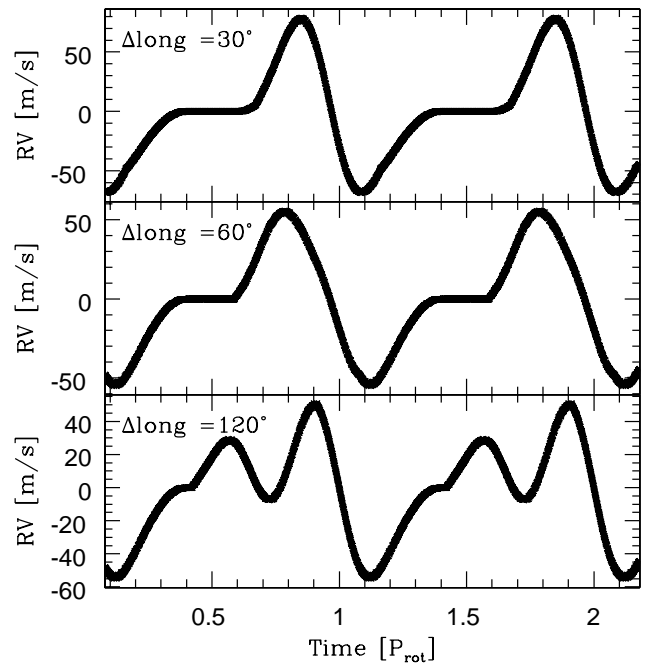}
\caption{RV as a function of time induced by two spots on the stellar surface. One spot is at latitude +30$^{\circ}$, the other at the equator. \textit{From top to bottom:} the shape of the variation varies with the difference in longitude between the two spots, labelled in the top left of each panel. }
\label{myfig13}
\end{center}
\end{figure}

\begin{figure}
\centering
\includegraphics[width=8.5cm]{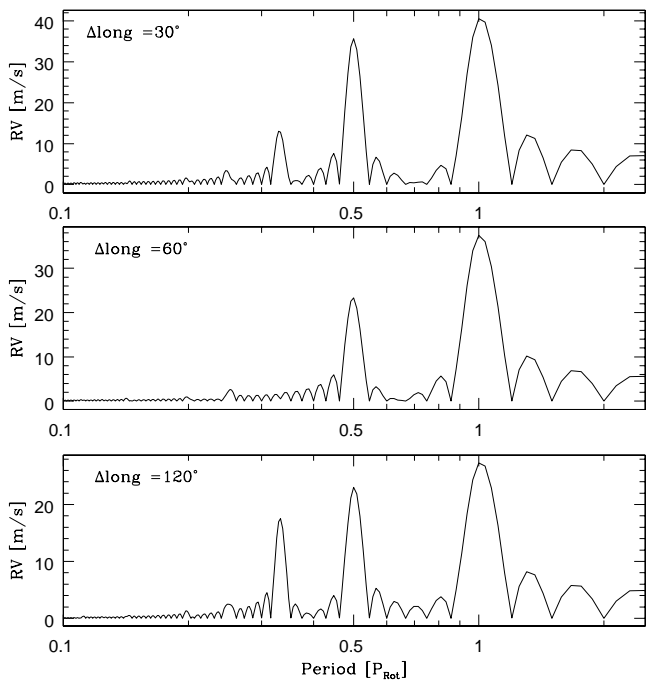}
      \caption{ Lomb-Scargle periodograms of the three RV variations showed in Fig.~\ref{myfig13} when two spots are on the stellar surface. Only the rotational period of the star and its two-first harmonics are detected. As noted in Sect.~\ref{RVspot}, the maximal peak is not at $P_{rot}/2$, so there is not a significant detection of the harmonic $P_{rot}/4$.
              }
         \label{myfig14}
   \end{figure}

	We simulated and studied the apparent shifts in RV induced by several spots on the stellar surface. Two spots of 1\% of the stellar surface are successively shifted by 30, 60, and 120$^{\circ}$ in longitude on an inclined star with $i$ = 40$^{\circ}$. One spot is fixed at latitude +30$^{\circ}$ and the other at the equator. Differential rotation is not considered. The three RV signatures and periodograms are plotted in Figs.~\ref{myfig13} and~\ref{myfig14}. Only the three periods $P_{rot}$, $P_{rot}/2$, and $P_{rot}/3$ are identified. Three sinusoids fixed at these periods fit the RV modulation and reduce the amplitude of the jitter by up to 90\%.
	

\begin{figure}
\begin{center}
\includegraphics[width=8.cm]{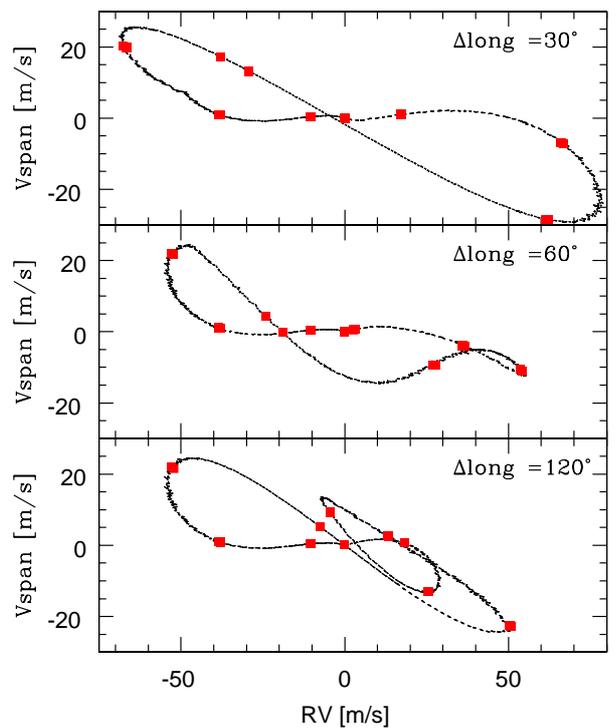}
\caption{V$_{span}$ as a function of RV for the three RV variations showed in Fig.~\ref{myfig13} when two spots are on the stellar surface. The square points represent a sampling with one measurement each P$_{rot}$/10. 
             }
\label{myfig15}
\end{center}
\end{figure}

For our three simulated cases with two spots, the anti-correlation between V$_{span}$ and RV is more or less observed on Fig.~\ref{myfig15} but may be affected by one spot possibly compensating for the other. We emphasized that the slope of the anti-correlation is shallow and requires data with a high sampling to be resolved (at least 10 measurements per rotation period).

\begin{figure}
\begin{center}
\includegraphics[width=8.cm]{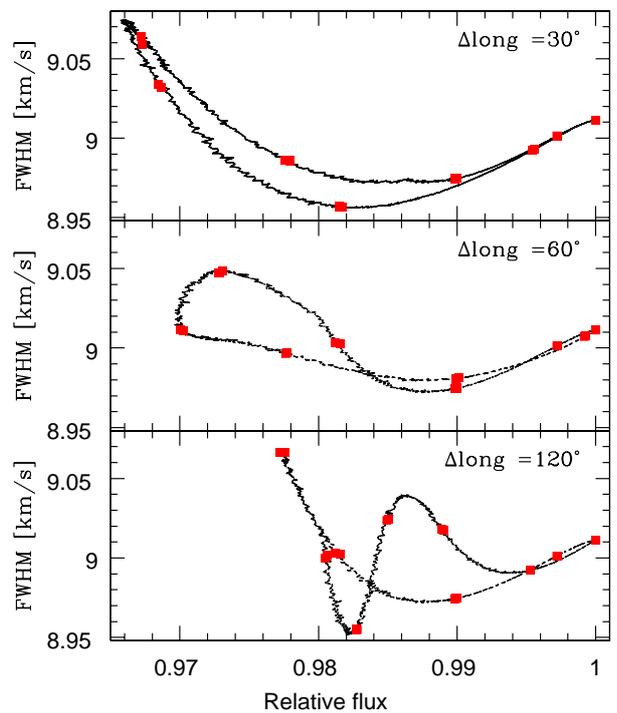}
\caption{FWHM of the fitted Gaussian as a function of the photometric flux of the three RV variations shown in Fig.~\ref{myfig13} when two spots are on the stellar surface. The square points represent a sampling with one measurement each P$_{rot}$/10.
           }
\label{myfig16}
\end{center}
\end{figure}

The pattern observed for one spot between the CCF FWHM and the photometry can be strongly modified by the presence of several spots as shown in Fig.~\ref{myfig16}. It may not display a clear signature and the anti-correlation may be barely seen.
However, as both parameters are not modified by the presence of a planet (except for a transit), this anti-correlation due to stellar activity would always be observable. This is in contrast to the RV and V$_{span}$ (or BIS) anti-correlation, which may be hidden by a planet in the RV data.
		
\subsection{Bright spots}

We consider only dark spots, but other inhomogeneities are present on the surface of an active star. For instance, plages are bright structures that may be related to dark spots in active regions. We simulated bright spots and characterized the induced variations in the RV, CCF parameters, and photometry. We found that the amplitudes of the variations are identical for characteristics similar to those of a dark spot (except the brightness), but the shapes of all parameters are reversed. The V$_{span}$ and the RV then remain anti-correlated. Similarly, the relation between parameters of the CCF and the photometry persists.

\section{Application to real data}

	\subsection{HD\,189733}
\begin{figure} 
\begin{center}
\includegraphics[width=8.4cm]{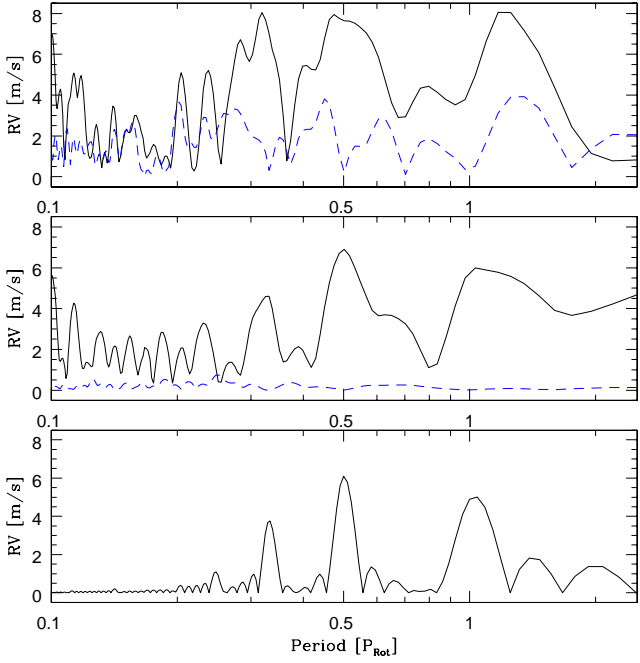}
\caption{\textit{Top:} Lomb-Scargle periodogram of the residuals from the Keplerian fit of the SOPHIE RV of HD\,189733 (black curve). The three peaks at the rotational period and the two first harmonics are detected. The Lomb-Scargle periodogram of the residuals after the simultaneous Keplerian fit and the harmonic filtering of the stellar activity is indicated by a dashed line. \textit{Middle:} Lomb-Scargle periodogram of a simulation of one dark spot on the surface of a star with parameters closest to those of HD\,189733 (black curve). The sampling is identical to that of the SOPHIE data for HD\,189733. The lomb-Scargle periodogram of the residuals after the simultaneous Keplerian fit and the harmonic filtering of the stellar activity is indicated by a dashed line. \textit{Bottom:} Lomb-Scargle periodogram of the same simulation as in the middle panel with a very good data sampling (one measurement each P$_{rot}$/1000). 
             }
\label{myfig17}
\end{center}
\end{figure}

	The active K2V star HD\,189733 and its transiting planetary companion, which has a 2.2-day orbital period, was monitored by Boisse et al. (2009). They used the high-resolution spectrograph \textit{SOPHIE}  mounted on the 1.93-m telescope at the Observatoire de
Haute-Provence to obtain 55 spectra of HD~189733 over nearly two months. The RV measurements subtracted from their fit of the planetary companion are variable because of the activity of the star. HD~189733 has a stellar rotational period of 11.953d (Henry \& Winn 2009). We computed the Lomb-Scargle periodogram of the residuals from the Keplerian fit in the top panel of Fig.~\ref{myfig17}. The fundamental period at P$_{rot}$ and its two first harmonics (P$_{rot}/2$ and P$_{rot}/3$) are detected with false alarm probability lower than $10^{-1}$. False alarm probabilities are computed by performing random permutations of the data, keeping the observing time fixed, as described in Lovis et al. (2010). There is no significant signal at other harmonics. 
The dispersion value of 9.1~ms$^{-1}$ are reduced to 5.5~ms$^{-1}$ when the residuals are fitted by three sinusoids fixed at the given periods. This dispersion is comparable to that obtained by the RV-BIS anti-correlation correction applied by Boisse et al. (2009). The main limitation here is the intrinsic precision of the spectrograph.
 The Lomb-Scargle periodogram of the residuals after the simultaneous fit of a Keplerian and three sinusoids with periods fixed at $P_{rot}$ and its two first harmonics is shown as a dashed line at the top of Fig.~\ref{myfig17}. It shows no evidence of another companion in the system taking into account the current SOPHIE accuracy ($\simeq$ 4-5ms$^{-1}$). 
	
	We show the effect of the sampling in the middle and bottom panels of the Fig.~\ref{myfig17}. We simulated a dark spot on the star with the closest parameters to those of HD~189733. Spot parameters (0.45\% of the visible stellar surface and $lat$=+30$^{\circ}$) were selected to ensure that the simulated photometric variations are comparable to those observed by MOST (Boisse et al. 2009). In the bottom panel, the Lomb-Scargle is computed for a quasi-perfect sampling during which a measurement is taken each P$_{rot}$/1000. In the middle panel, the sampling is that of the measurements of HD~189733 in Boisse et al. (2009). The pattern of the Lomb-Scargle periodogram for the simulations with the sampling of the SOPHIE data (\textit{middle panel}) is comparable to those of real data  (\textit{top panel}). As a dashed line, we superimpose the Lomb-Scargle periodogram of the residuals of the harmonic filtering of the simulated RV with the SOPHIE data sampling. No periodicities are found to have greater amplitudes because of the harmonic filtering beyond the noise level.

	\subsection{GJ\,674}
\begin{figure}[h] 
\begin{center}
\includegraphics[width=4.45cm]{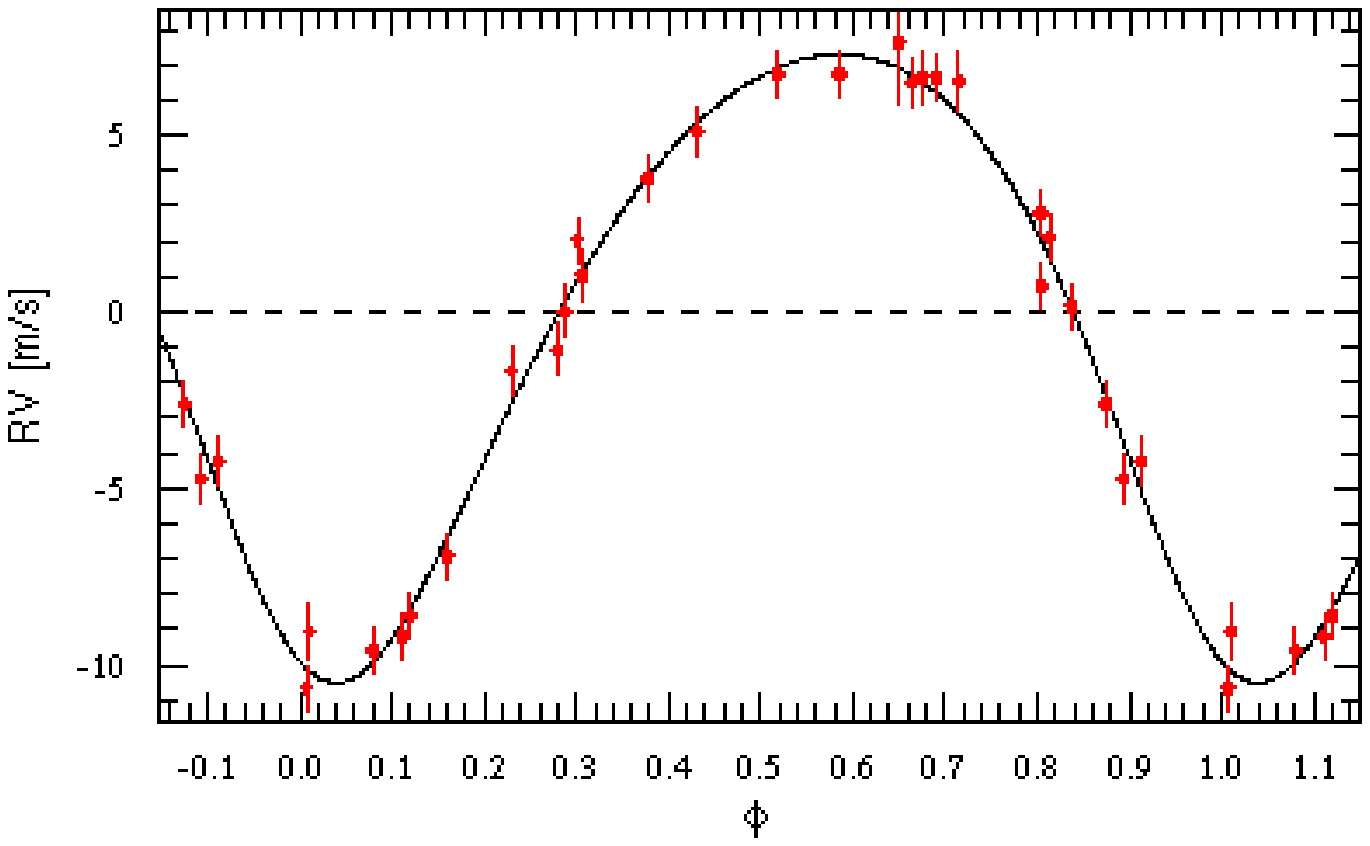}
\includegraphics[width=4.45cm]{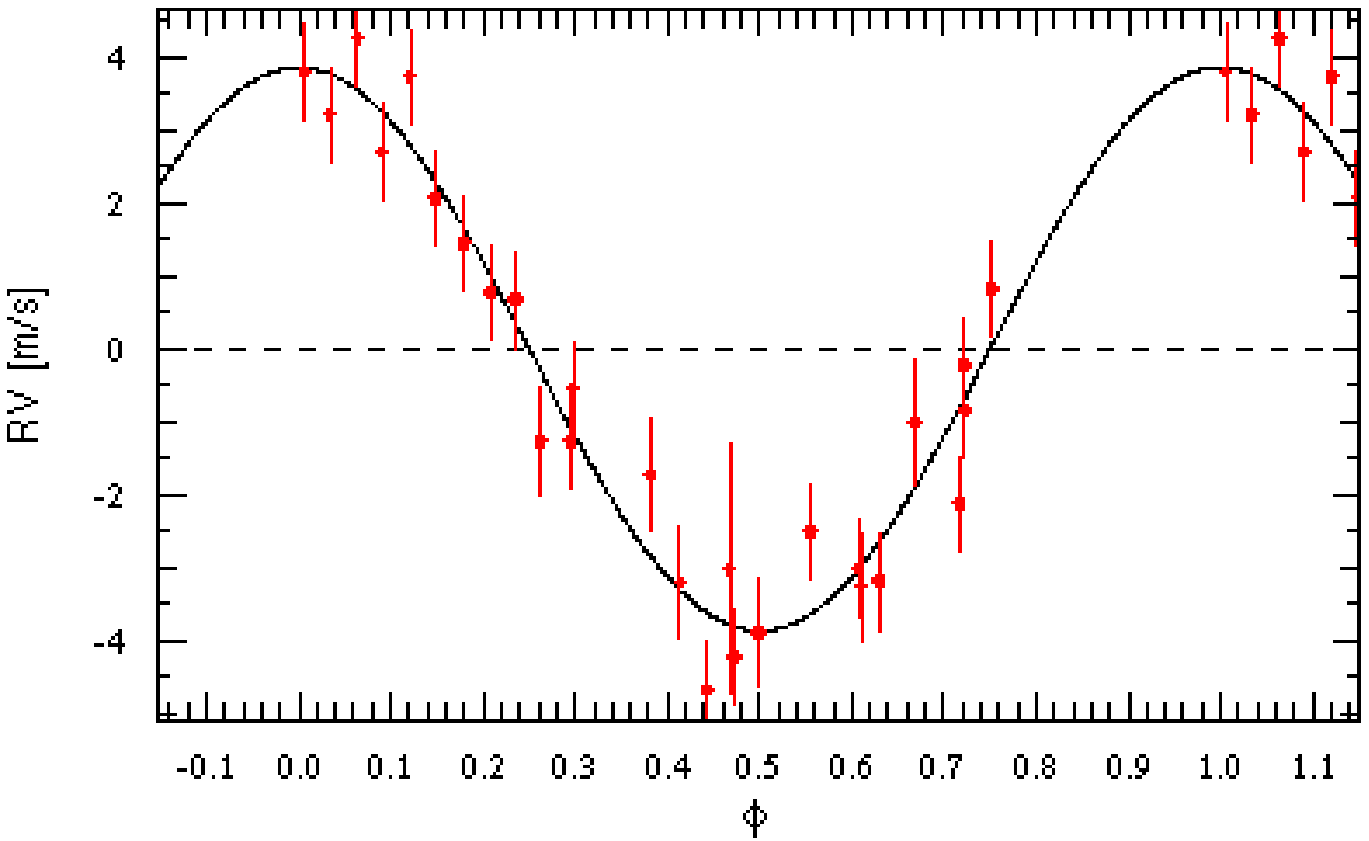}
\includegraphics[width=4.45cm]{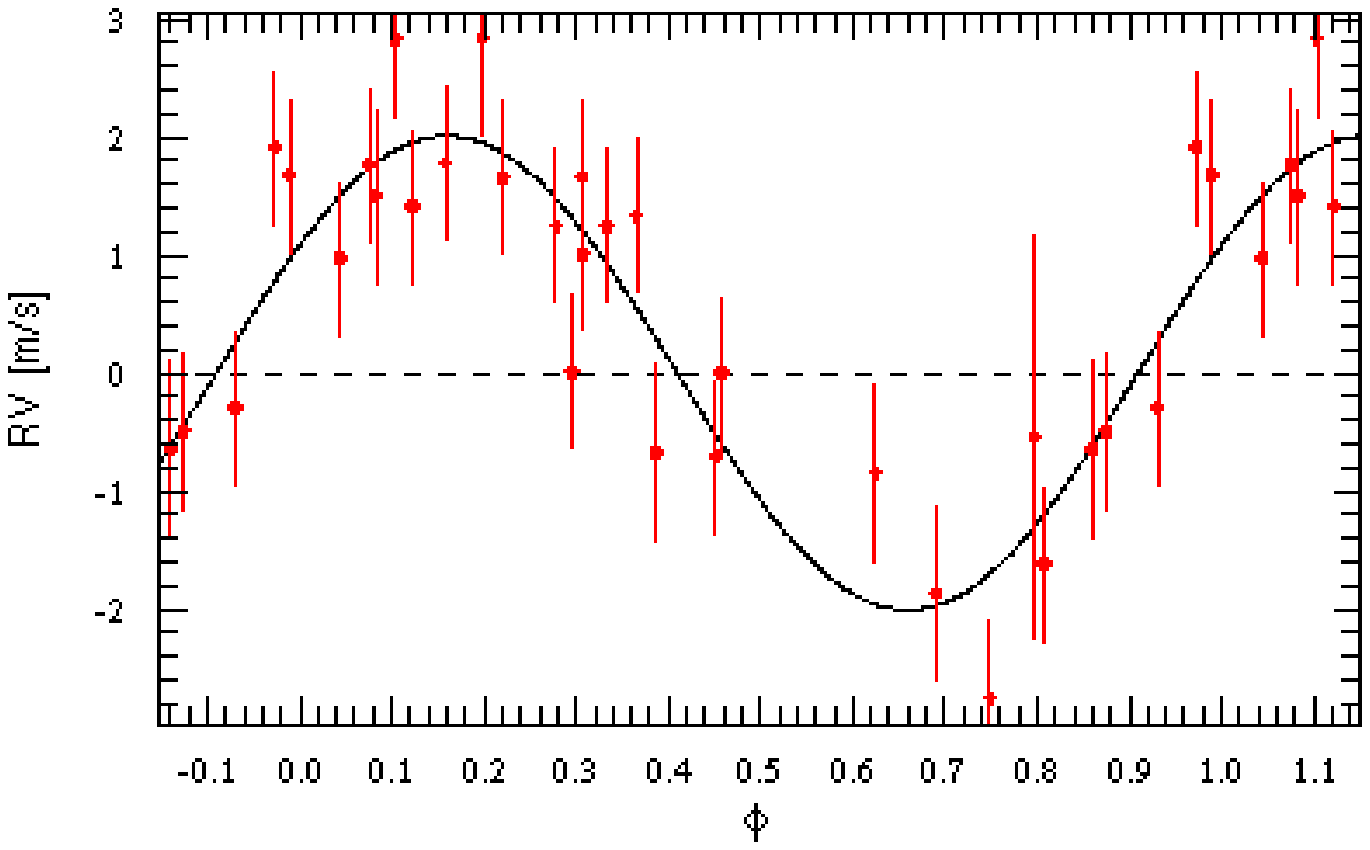}
\includegraphics[width=4.45cm]{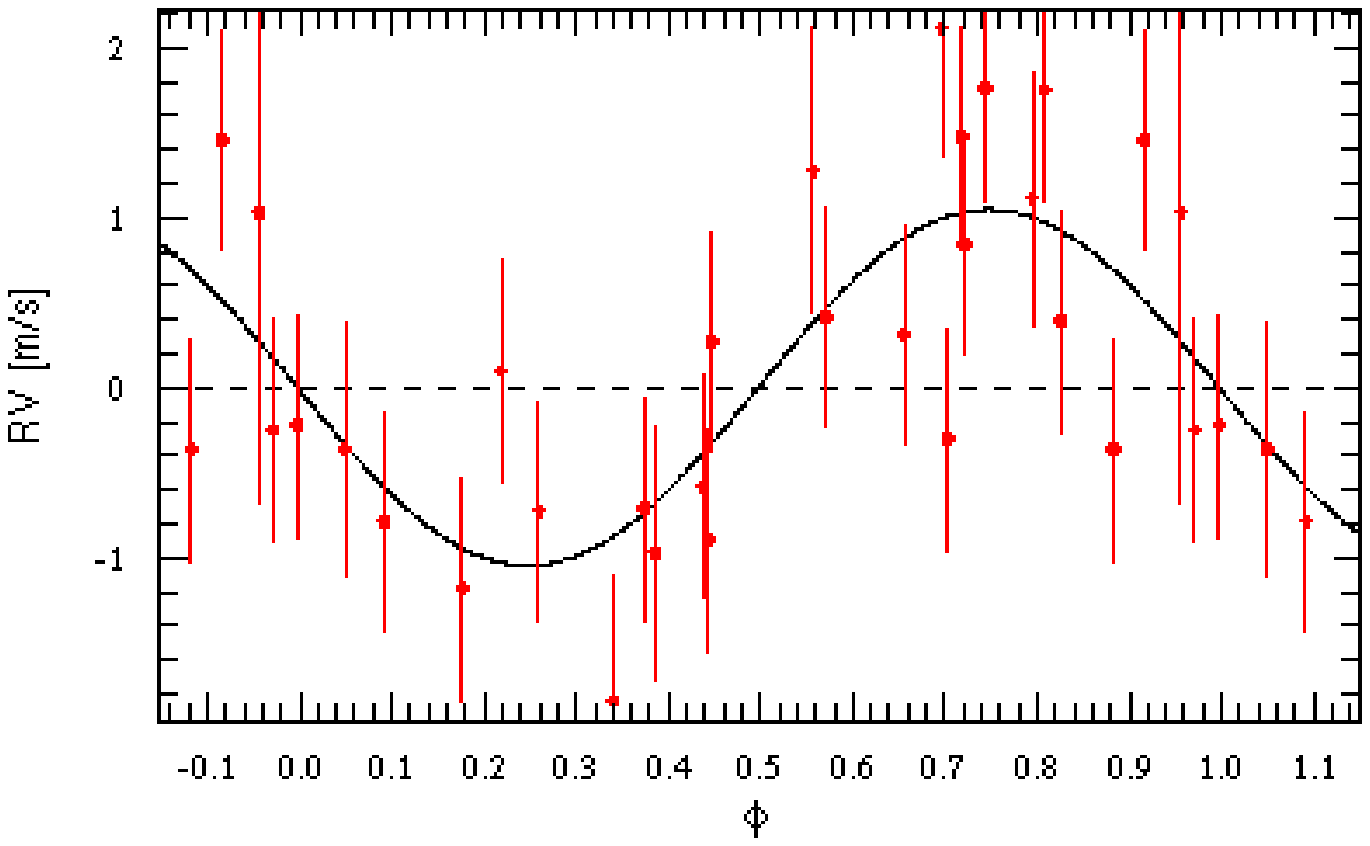}
\caption{Fit of the HARPS RV of GJ674 with one Keplerian and three sinusoids fixed to the rotational period (\textit{top right} P$_{rot}$=34.85d) and the two first harmonics (\textit{bottom left} P$_{rot}/2$=17.425d and \textit{bottom right} P$_{rot}/3$=11.6167d). The Keplerian fit (\textit{top left}) gives the planetary parameters. Each plot is given as a function of orbital phase after removing the other three signals.
Individual error bars are also plotted.}
\label{myfig18}
\end{center}
\end{figure}

	GJ\,674 is a moderately active M2.5V dwarf hosting a planet with a 4.69-day period (Bonfils et al. 2007). A superimposed signal with a periodicity of roughly 35 days is also visible in the RV measurements. Bonfils et al. (2007) demonstrated that their variations coincide with the stellar rotation period by analyzing photometry and active lines (\ion{Ca}{II} and \ion{H}{$\alpha$}). These signals originate in active regions rotating with the stellar surface. They described the data with a two-Keplerian model with a 4.69-day period planet and a Keplerian to fit the active feature.  
In Fig.~\ref{myfig18}, we fit the RV derived from HARPS spectra with three sinusoids each of period fixed to be the rotational period (P$_{rot}$=34.85d) and the two first harmonics (P$_{rot}/2$=17.425d and P$_{rot}/3$=11.6167d) and one Keplerian that provides the planetary parameters. These fitting results are in agreement with Bonfils et al. (2007) (cf. Table~3), being no significant change in either the fitted parameters or the error bars. On the other hand, we obtained a weaker dispersion in the residuals, $\sigma(O-C)$=0.65~ms$^{-1}$ instead of $\sigma(O-C)$=0.82~ms$^{-1}$, which is closest to the current HARPS accuracy and equal to the uncertainty in each measurement, and we were able to reduce the $\chi^{2}$ to 1.36.

\begin{figure}[h] 
\begin{center}
\includegraphics[width=8.5cm]{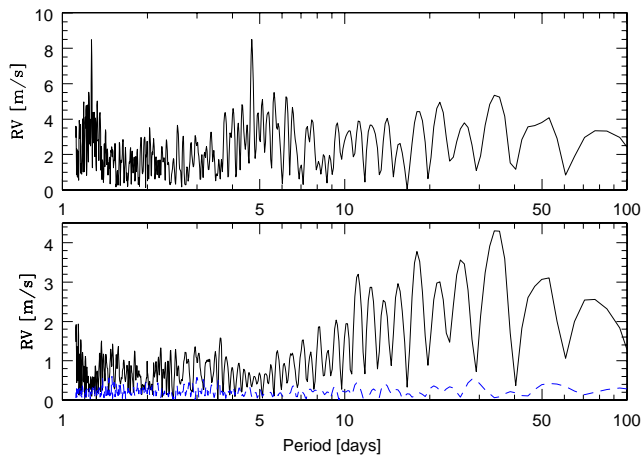}
\caption{\textit{Top:} Lomb-Scargle periodogram of the RV of GJ674. The main peaks are due to the periodicity of the planet and its one-day alias. \textit{Bottom:} Lomb-Scargle periodogram of the residuals from the Keplerian fit of the GJ674 RV data (black curve). The main peak is the rotational period of the star. The two first harmonics are also visible. The Lomb-Scargle periodogram of the residuals after the simultaneous Keplerian fit and harmonic filtering of the stellar activity is shown as a dashed line. }
\label{myfig18bis}
\end{center}
\end{figure}

For the original RV data, we computed the Lomb-Scargle periodogram, (Fig.~\ref{myfig18bis}, \textit{top}), which shows that the main period is that of the planet. After removing the RV variations due to the planet, the periodogram of the residuals displays periodicities at the rotational period and its two first harmonics (Fig.~\ref{myfig18bis}, \textit{bottom}). After the simultaneous fit of a Keplerian and the harmonic filtering, the periodogram shows neither a residual nor an enhanced periodicity.
	 
\begin{table}[h]
  \caption{Keplerian orbital solutions for GJ674b planet as published in Bonfils et al. (2007) and fitted in this paper.}
  \label{param_p}
\begin{tabular}{lcc}
\hline
\hline
Planetary parameters &   Bonfils et al. (2007) & This paper \\
\hline
$P_{P}$    [days]   &   4.694 $\pm$ 0.007     & 4.694 $\pm$ 0.002  \\
K          [\,m\,s$^{-1}$]           &  8.70 $\pm$ 0.19           & 8.9 $\pm$ 0.3 \\
e                              & 0.20 $\pm$ 0.02            & 0.19 $\pm$ 0.03 \\
$\omega$    [deg]  & 143 $\pm$  6                  & 159 $\pm$ 10 \\
T$_{0}$    [JD]        & 53780.09 $\pm$  0.08  & 53780.25 $\pm$ 0.12 \\
\hline
$m_{2}$$\sin$$i$$^{\mathrm{a}}$ [$M_{\oplus}$] & 11.09  & 11.39\\
$\sigma_{(O-C)}$$^{\mathrm{b}}$      [\,m\,s$^{-1}$]       & 0.82  & 0.65\\
reduced $\chi^{2}$ & 2.57 & 1.36 \\
\hline
\end{tabular}

$^{\mathrm{a}}$ assuming M$_{\star}$ = 0.35 M$_{\odot}$ (Bonfils et al. 2007)\\
$^{\mathrm{b}}$ $\sigma_{(O-C)}$  after the fit\\
\end{table}

     \subsection{CoRoT-7}
The photometric transit search with the CoRoT satellite has reported the discovery of a planetary companion CoRoT-7b around an active V=11.7 G9V star (L\'eger et al. 2009) with an orbital period of 0.85 days. Queloz et al. (2009) reported on the intensive campaign carried out with HARPS at the 3.6-m telescope at La Silla. The RV variations are found to be dominated by the activity of the star, i.e. rotational modulation from cool spots on the stellar surface 
with an estimated period close to 23 days. Two approaches were used to disentangle the Doppler motion from the active jitter. They each detected two signals: the CoRoT-7b transit period and a second planet CoRoT-7c at a period of 3.69 days. The authors inferred a mass of 4.8 $\pm$ 0.8 M$_{Earth}$ for CoRoT-7b and, assuming that both planets are on coplanar orbits, a mass of 8.4 $\pm$ 0.9 M$_{Earth}$ for CoRoT-7c. 
The second of the two approaches used in Queloz et al. (2009) is a modeling of the active jitter by a harmonic decomposition of the rotational period. The authors subtracted their model from the RV data before detecting and characterizing the planetary system. Here, we wish to fit simultaneously the effect of both activity and the planetary system on the data as we did previously for GJ674.

\begin{figure}[h] 
\begin{center}
\includegraphics[width=8.5cm]{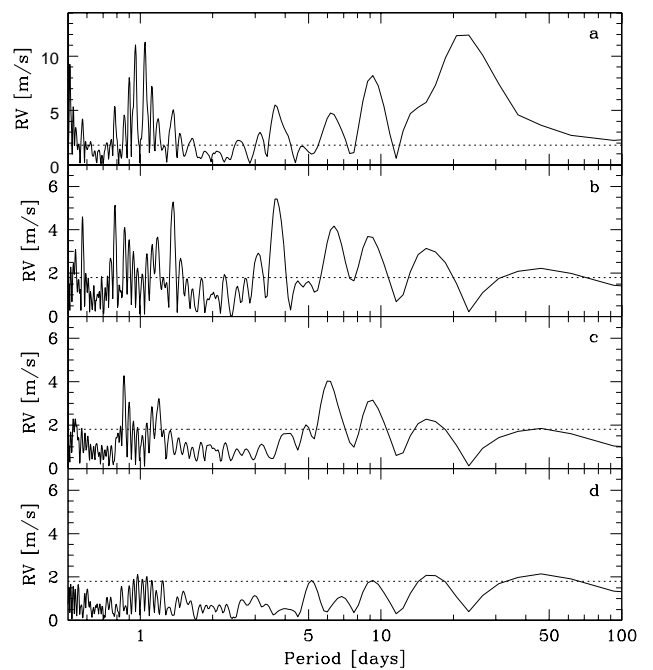}
\caption{Lomb-Scargle periodograms of the 37 last days of HARPS RV data of CoRoT-7. The dotted line shows the level of the mean uncertainty in RV measurements. a) The main peak in the raw data is owing to the rotational period of the star ($\sim 23d.$). b) After the harmonic filtering of the data, the highest peak of the residuals corresponds to the CoRoT-7c period. c) The residuals of the simultaneous fit of the activity and one Keplerian for CoRoT-7c show two peaks, the highest at the CoRoT-7b period and the other at its one-day alias. d) No significant periods remains above the level of the noise after the simultaneous fit of the activity and of two Keplerians for the possible companions.}
\label{myfig18ter}
\end{center}
\end{figure}

We used the 37 last days of HARPS data when the sampling had a higher frequency on a long timescale (several points each night and a total of 59 measurements) and in order that the distribution of spots on the stellar surface did not change too much (and therefore the parameters of the fit for the active signal). This assumption is consistent with that of the photospheric activity study of Lanza et al. (2010) for the CoRoT photometric measurements. They found that active regions occur on three stable active longitudes, whose overall lifetime may exceed the duration of the light curve. They note that the large active regions evolve on timescales ranging from two weeks to a few months. However, we show in Sect. 2 that an evolving active region is also closely fitted by our harmonic filtering. We fitted simultaneously three sinusoids to the active jitter with periods fixed at the rotational periods (23d.) and its two first harmonics (11.5d and 7.66667d.). The Lomb-Scargle of the residuals shows a clear peak near 3.69\,d and another one near 0.85\,d with false alarm probabilities lower than $5.10^{-4}$ (Fig.~\ref{myfig18ter}). No other significant periods are detected with false alarm probability greater than 0.5. We fitted simultaneously three sinusoids to the active jitter and two Keplerians to the possible companions. No parameters were fixed for the Keplerians except the eccentricities, which were fixed to be zero. The differences from the published values of the periods are smaller 0.5\% and of the transit phase of CoRoT-7b is less than 0.2\% of the transit period. To measure the semi-amplitude and then the mass of the planets, we fixed the period and the T0 of the transiting companion. We then fitted simultaneously three sinusoids for the active jitter and two Keplerians with null-eccentricity. We obtained the parameters given in Table~\ref{param_c}. The period of 3.70 $\pm$ 0.02 d found for CoRoT-7c agrees with the value of 3.698 $\pm$ 0.003 d found by Queloz et al. (2009). The residuals of 3.1 ms$^{-1}$ and the reduced $\chi^{2}$, which is equal to 2., are comparable to those obtained by Queloz et al. (2009) considering that we have used fewer data. We note, however, that the semi-amplitude of the planets 4.5 $\pm$ 0.7 ms$^{-1}$ for CoRoT-7b and 6.1 $\pm$ 0.6 ms$^{-1}$ for CoRoT-7c are larger than in Queloz et al. (2009). For comparison, the same study is performed on another data set of bjd=[54775.8-54807.3], with a total of 34 measurements. The two planets are detected in the Lomb-Scargle periodogram with false alarm probabilities lower than $10^{-2}$, whereas no other periods are detected with false alarm probability lower than 0.5 (except for the one-day alias due to the data sampling). 
The differences from the published values are 4.5\% for the CoRoT-7c period, 2\% for the CoRoT-7b transit period, and 9\% for the CoRoT-7b transit phase. The semi-amplitude for the planets are 2.9 $\pm$ 1.1 ms$^{-1}$ for CoRoT-7b and 4.6 $\pm$ 0.9 ms$^{-1}$ for CoRoT-7c. 

Our method is robust, but these differences illustrate the difficulty in measuring the amplitudes accurately in the presence of activity. 
Moreover, it led us to conclude that the error bars for the semi-amplitudes are underestimated in our study as in that of Queloz et al. (2009). 
We infer approximately that a systematic noise due to active jitter of 1.5\,ms$^{-1}$ must be added quadratically to the error bars. We then determine a semi-amplitude of 4.0\,$\pm$\,1.6\,ms$^{-1}$ for CoRoT-7b and 5.6\,$\pm$\,1.6\,ms$^{-1}$ for CoRoT-7c. 
The masses are 5.7\,$\pm$\,2.5 M$_{Earth}$ for CoRoT-7b, which agrees with the value of Queloz et al. (2009), and 13.2\,$\pm$\,4.1 M$_{Earth}$ for CoRoT-7c (assuming that both planets are on coplanar orbits), slightly higher than its published value. The error bars then account for a systematic noise due to active jitter and the stellar mass uncertainty (0.93\,$\pm$\,0.03\,M$_{\odot}$). 
By evaluating a weighted average, we find for CoRoT-7c a period of 3.697 $\pm$ 0.019 day, in agreement with Queloz et al. (2009). These new values for the CoRoT-7 system are reported in Table~\ref{param_c2}. Considering our value of the mass of CoRoT-7b, one sigma higher than the published value, and its refined measurement of radius, slightly smaller $R_{P}$ = 1.58 $\pm$ 0.10 R$_{Earth}$, derived by Bruntt et al. (2010) using improved stellar parameters, Valencia et al. (2010) predicted that is a planet compatible with an Earth-like composition (33\% iron, 66\% silicate).

\begin{table*}[t]
\caption{CoRoT-7 fit with three sinusoids for the active jitter and two Keplerians for the planets for the last 37 days of HARPS RV measurements.}
 \label{param_c}
\begin{tabular}{cccccc}
\hline
\hline
Parameters&&& \\
\hline
$P_{P}$ [days]   &   $P_{rot}$=23 (fixed) & $P_{rot}/2$=11.5 (fixed) & $P_{rot}/3$=7.66667 (fixed) & 3.695 $\pm$ 0.02 & 0.8536 (fixed) \\
K  [\,m\,s$^{-1}$]     &  14.3 $\pm$ 0.8 & 3.7 $\pm$ 0.8 & 1.2 $\pm$ 0.2 & 6.1 $\pm$ 0.6 & 4.5 $\pm$ 0.7 \\
e             &     &   &  & 0 (fixed) & 0 (fixed)\\
T$_{0}$  [JD]  &  &   &  & 54899.2 $\pm$ 0.7 & 54899.761 (fixed) \\ 
$\sigma_{(O-C)}$ [\,m\,s$^{-1}$]  &    & before the fit 10. & after the fit 3.1; reduced$\chi^{2}$ = 2.\\
\hline
\end{tabular}
\end{table*}

\begin{table}[t]
  \caption{Adopted values for the two CoRoT-7 planets as derived from this paper.}
  \label{param_c2}
\begin{tabular}{ccc}
\hline
\hline
Parameters & CoRoT-7b & CoRoT-7c  \\
\hline
$P_{P}$ [days]   &  0.8536 (fixed) & 3.697 $\pm$ 0.019 \\
K  [\,m\,s$^{-1}$]   & 4.0 $\pm$ 1.6$^{\mathrm{\dag}}$   &  5.6 $\pm$ 1.6$^{\mathrm{\dag}}$ \\
e                           & 0 (fixed) & 0 (fixed)\\
T$_{0}$  [JD]  & 54899.761 (fixed)  & 54899.2 $\pm$ 0.7 \\ 
$m_{2}$ [$M_{\oplus}$] & 5.7 $\pm$ 2.5$^{\mathrm{\ast}}$  & 13.2 $\pm$ 4.1$^{\mathrm{\ast}}$\\
\hline
\end{tabular}

$^{\mathrm{\dag}}$ Error bars in semi-amplitudes take into account a systematic noise due to active jitter of 1.5\,m\,s$^{-1}$.\\ 
$^{\mathrm{\ast}}$ Error bars in masses take into account the stellar mass uncertainty and a systematic noise due to active jitter.\\
\end{table}

	\subsection{$\iota$ Hor}
		$\iota$\,Hor, or HD\,17051, is a young G0V star with V~=~5.40. The 320.1-d period planet $\iota$\,Hor\,b with a small eccentricity of 0.161 was reported by  K\"urster et al. (2000).
They noted an excess RV scatter of 27 ms$^{-1}$ due to stellar activity. Rocha-Pinto \& Maciel (1998) measured a \ion{Ca}{II} index of log$R^{'}_{HK}$=-4.65, and Jenkins et al. (2006) measured $\log$$R^{'}_{HK}$=-4.59. Asteroseismologic observations were made with the high-precision spectrograph HARPS of $\iota$\,Hor (Vauclair et al. 2008). They studied the acoustic oscillations of the star and demonstrated that $\iota$\,Hor was formed in the same primordial cloud as the Hyades. The data helped refine the mass of the star $M_{*}$\,=\,1.25\,$\pm$\,0.01\,M$_{\odot}$, which lead us to re-evaluate the $\iota$\,Hor\,b planetary mass to be 2.6 M$_{Jup}$.
			
	We studied these data to characterize the active jitter and search for a possible hidden Doppler motion. Observations were carried out during eight consecutive nights between 19 and 26 November 2006 with 1856 measurements each of time exposure of about 100s. The p-modes are visible at high frequencies and a low frequency RV signal with amplitude of about 20~ms$^{-1}$ appears because of the activity. We kept 1659 measurements with SNR(550nm)~$\geqslant$~120. We averaged the data in groups of 20 measurements in order to average the p-mode signature. The mean RV photon-noise uncertainty in the averaged points is then about 26~cms$^{-1}$, but the true precision is limited by the instrumental accuracy $\approx$ 80\,cms$^{-1}$. 
		
	Before studying the RV variations due to stellar activity, we subtracted the long-period planet Doppler motion. We computed the $\iota$\,Hor\,b Keplerian solution with the parameters of K\"urster et al. (2000) because our data has a too short duration and cannot be used to better constrain the planetary orbit. We subtracted a slope of -650~ms$^{-1}$yr$^{-1}$ from the RV measurements. The uncertainty in the K\"urster et al. (2000) parameters, particularly in the period ($\delta P$ = 2.1d) and the epoch of maximum RV ($\delta T_{0}$ = 3.0 d), induced an uncertainty of $\pm 50$~ms$^{-1}$yr$^{-1}$ in the value of the slope at the epoch of our observations, approximately eight years after the Kurster ones. \\ 
	
\begin{figure}
   \centering
   \includegraphics[width=8.7cm]{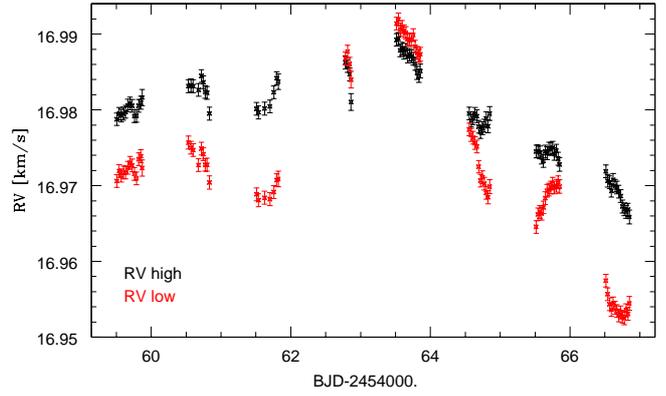}
      \caption{ RV$_{high}$ and RV$_{low}$ of $\iota$\,Hor derived from HARPS spectra as a function of time. The slope due to the known planet has not been removed. The error bars are also plotted.            }
         \label{myfig19}
   \end{figure}

\begin{figure}
   \centering
   \includegraphics[width=7.5cm]{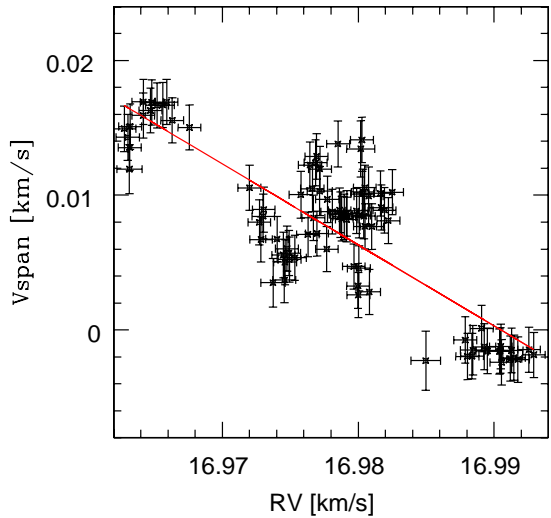}
      \caption{ V$_{span}$=RV$_{high}$ - RV$_{low}$ as a function of RV of $\iota$\,Hor derived from HARPS spectra. The line is the least squares fit. The ranges have the same extents along the x- and y-axes. The error bars are also plotted. One may compare this shape with that of the simulation of two spots separated by 120$^{\circ}$ in longitude in Fig.~\ref{myfig15} (\textit{bottom}). }
         \label{myfig20}
   \end{figure}

		To check that the main RV variations is due to activity, we computed and compared RV$_{high}$ and RV$_{low}$ in Fig.~\ref{myfig19}. The amplitude of RV$_{low}$ is roughly 1.5 times the amplitude of RV$_{high}$, as expected in the case of stellar activity (cf. Sect.~\ref{ccfparameters}). Thus, in Fig.~\ref{myfig20} an anti-correlation is observed between the RV and the V$_{span}$ in agreement with an active signature. 

		
		 The study of the dark spot simulations in Sect.~\ref{simulations} shows that the active RV jitter is well-fitted when the rotational period of the star is known. The $\iota$\,Hor rotational period was estimated by Saar \& Osten (1997) and by Saar et al. (1997) from \ion{Ca}{II} emission. They found $P_{rot}$ = 7.9d and 8.6d, respectively.  From the Kurucz model atmosphere, K\"urster et al. (2000) derived a $v$\,$\sin$\,$i$ = 5.5\ kms$^{-1}$, which agrees with the 5.7~kms$^{-1}$ found by Saar \& Osten (1997). Also in agreement, we derived from the mean FWHM of the HARPS CCF a value of $v$\,$\sin$\,$i$ = 5.7~kms$^{-1}$. Using the Saar \& Osten (1997) formula to estimate the $\sin$\,$i$ from the $v$\,$\sin$\,$i$, $P_{rot}$, and the stellar radius, we derived the stellar inclination $i$=55$\pm$10$^{\circ}$. Following Saar \& Donahue (1997), we estimate that the $f_{s}$, nonuniform portion of the spot distribution responsible for the observed variability is $f_{s}\approx$0.3\%.
   \begin{figure}
   \centering
   \includegraphics[width=8cm]{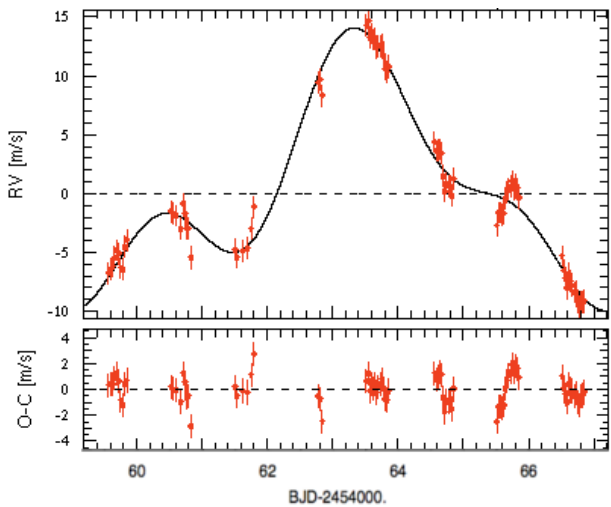}
      \caption{ \textit{Top:} RV of $\iota$ Hor derived from HARPS spectra as a function of time. The  three-sinusoid fit, with periods fixed at the rotational period of the star and its two first harmonics, is plotted as a solid line. The rotational period is chosen to be equal to 8.2 days. \textit{Bottom:} Residuals from the fit as a function of time. The dispersion is equal to 1.03~ms$^{-1}$. 
              }
         \label{myfig22}
   \end{figure}
   \begin{figure}
   \centering
   \includegraphics[width=7.5cm]{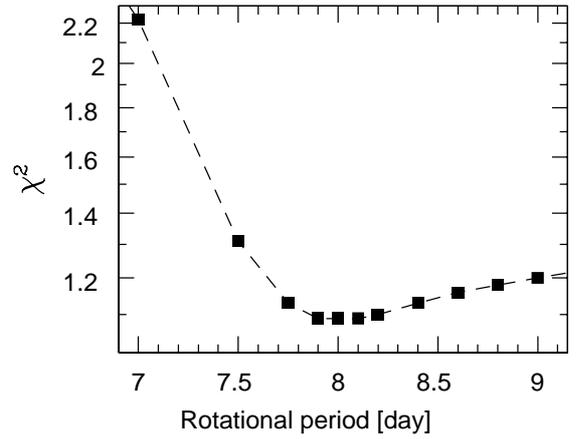}
      \caption{ Reduced $\chi^{2}$ as a function of the value of the rotational period used for the fit of the $\iota$\,Hor data. The rotational period is used to fix the period of the fitted sinusoids. Minimum values of the $\chi^{2}$ are obtained for $P_{rot}$ between 7.9d and 8.4d. 
            }
         \label{myfig21}
   \end{figure}

We choose arbitrarily a rotational period equal to 8.2 days. The RV are fitted by three sinusoids with periods fixed at the rotational period and its two first harmonics (cf. Fig.~\ref{myfig22}). The residuals are equal to $\sigma$~=~1.03ms$^{-1}$ reaching almost the instrumental accuracy. 
A peak at the third harmonic is barely detected above the noise in the Lomb-Scargle periodogram of the residuals, as expected when data still exhibit signature of activity (cf. Sect.~\ref{rvfitofadarkspot}). Fitting simultaneously the fourth sinusoids, the semi-amplitude of the fourth one is only 1.5 ms$^{-1}$ and the residuals are slightly smaller at a value of $\sigma$~=0.87ms$^{-1}$. 
The amplitude of this last sinusoid represents the limit of our detection ($\lesssim$\,2\,$\sigma$) as before in the study of the dark spot simulations, illustrating that three sinusoids is sufficient to remove most of the active jitter.
In Fig.~\ref{myfig21}, it is shown that minimum values of $\chi^{2}$ are obtained for $P_{rot}$ between 7.9d and 8.4d.

\begin{figure}
   \centering
   \includegraphics[width=8.5cm]{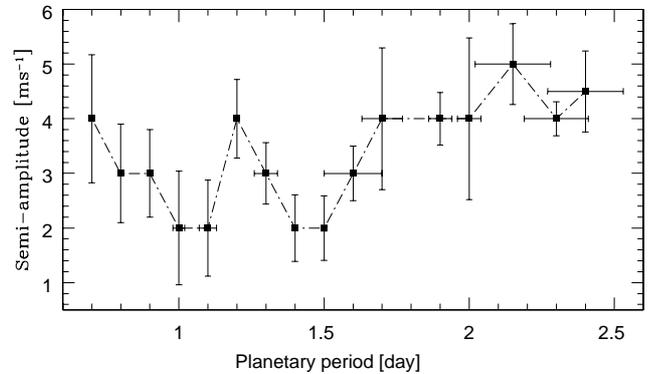}
      \caption{ Minimal semi-amplitude of a planet that would be detected in the $\iota$Hor data as a function of the planetary period. The error bars represent the difference between the fitted and the simulated parameters, or, if it is greater the error in the fitted parameters.
               }
         \label{myfig19bis}
   \end{figure}

In the $\iota$Hor data, we do not detect a short-period companion. Nevertheless, we wished to check whether we have subtracted the RV shift due to a companion when subtracting the effects of activity. We ran simulations and added RV due to fake planets to the $\iota$Hor data. We consider only the case of circular orbits, which is a good hypothesis for planets of period shorter than 6 days (Eggenberger \& Udry 2010). We fitted the active jitter with three sinusoids of period fixed to the rotational period and its two first harmonics. Afterwards, we considered the Lomb-Scargle periodogram of the residuals. If a peak at the planetary period was detected, we fitted simultaneously a Keplerian with null-eccentricity to obtain the planetary parameters. We consider that a peak is significant if its false-alarm probability is smaller than 10$^{-2}$. In Fig.~\ref{myfig19bis}, we plotted the minimal semi-amplitude detected in the $\iota$Hor data as a function of the planetary period. The error bars represent the difference between the planetary period and semi-amplitude that were fitted and the simulated ones or, if it is greater, the error in the fitted parameters. Periods longer than 7 days were not considered because observations during at least one planetary period were needed to detect a planet. Planetary periods between 7 and 2.5 days are not easily characterized because, without other high-precision measurements, we cannot evaluate the amplitude of the RV jitter due to activity and then, estimate above which amplitude the RV variation might be due to a companion. When shorter than 2.5 days, the periods are detected in the periodogram of the residuals, after subtraction of the RV jitter.

To conclude, the $\iota$Hor RV modulations can be clearly explained by dark spots on the photosphere of a star with a rotational period in the range [7.9-8.4] days. There is no doubt that the main RV variations are due to the stellar activity by the anti-correlation RV-V$_{span}$ (Fig.~\ref{myfig20}) and the difference in amplitude between the RV$_{low}$ and RV$_{high}$ (Fig.~\ref{myfig19}). 
We excluded the presence of planets with minimum masses between 6 and 10 M$_{Earth}$ and respective periods  between 0.7 and 2.4 days (Fig.~\ref{myfig19bis}). 
The causes limiting our diagnostic are primarily the uncertainty in the rotational period of the star, the short observational period, and, to a lesser extent, the uncertainty in the slope due to the $\iota$Hor\,b planet.

\section{Conclusion and Perspectives}

We have performed simulations of stellar dark spots to check the validity of the standard
diagnostics of activity and to derive a method to subtract RV variations due to activity.
We have studied the effects of stellar spots on photometry, measured radial velocities, and other
spectral parameters, including the asymmetry of the mean line of the spectrum that we
estimate with a new parameter, V$_{span}$=RV$_{high}$ - RV$_{low}$, which is less sensitive to noise than the usual bisector span BIS. 
We have shown that the anti-correlation between RV and the V$_{span}$ needs a minimal
sampling of 0.1 of a stellar rotational period to keep its diagnostic power for several spots,
and that an anti-correlation might be observed between the photometry
and the FWHM of the spectral lines.
We have pointed out that bright spots also induce an anti-correlation between RV and the V$_{span}$.
We have shown that stellar spots induce RV variations at the rotational period $P_{rot}$ of
the star and its two first harmonics $P_{rot}/2$ and $P_{rot}/3$. These RV variations
can be corrected by three sinusoids, the remaining activity signal at the following
harmonic representing less than 10~\%\ of the initial RV amplitude.
Fitting stellar activity with three sinusoids may be used to disentangle it from Doppler motion
provided that 1) the period of the planet is not close to that of the stellar rotation or one of
its first harmonics, and 2) the rotational period of the star is accurately known or the data
cover at least two rotational periods of the star. Under these conditions, this method
could even reveal low-mass planets with semi-amplitudes down to about 30\% of the semi-amplitude of the activity-induced variation.

These results have been validated on four known active planet-host stars.
In the case of CoRoT-7, we have also detected the two planets reported by
Queloz et al. (2009) but our uncertainties in the fitted parameters are several times larger to properly take into account  the stellar active jitter. Moreover, with our simultaneous modeling of the activity and planetary parameters, slightly higher masses are found: 
5.7 $\pm$ 2.5 M$_{Earth}$ for CoRoT-7b and 13.1 $\pm$ 4.1 M$_{Earth}$ for CoRoT-7c (assuming that both planets are on coplanar orbits). In the case of $\iota$\,Hor, we have been able to exclude
low-mass planets with periods between 0.7 and 2.4 days with semi-amplitude
greater than 4 ms$^{-1}$, which correspond respectively to a minimum mass of between 6 and 10 M$_{Earth}$. 

Our simulations did not account for the differential rotation. This
may generate frequencies other than the harmonics of the rotational
period. These additional periodicities, however, were not detected in our study of known
active stars. On the other hand, for an RV survey, when a periodicity close to the
rotational period of the star or its harmonics is found, a photometric follow-up has to be
performed.
As illustrated in the  discovery and confirmation of the planetary
companion of HD\,192263, a planet may be found with a period roughly close to that of the
photometric variations of the star (Santos et al. 2000; Henry et al. 2002;
Santos et al. 2003). Simultaneous photometry with the RV is then needed to
check the phase of the parameters.

This method for distinguishing stellar activity and planetary signals is
particularly suitable for RV follow-up of transit surveys from the ground or space such as CoRoT
or Kepler.
In this case, photometry of active stars is indeed well sampled and
characterized (Paulson et al. 2004). The method could also be used for RV survey around (low)
active stars when the companion candidate has a periodicity that differs from the
rotational period of the star and its harmonics, to refine the planetary parameters.

\begin{acknowledgements}
     We wish to thank the French National Research Agency (ANR-08-JCJC-0102-01) for their support.  NCS would like to thank the support by the European Research Council/European Community under the FP7 through a Starting Grant, as well from Funda\c{c}\~ao para a Ci\^encia e a Tecnologia (FCT), Portugal, through a Ci\^encia\,2007 contract funded by FCT/MCTES (Portugal) and POPH/FSE (EC), and in the form of grants reference PTDC/CTE-AST/098528/2008 and PTDC/CTE-AST/098604/2008. The authors thanks the referee for his careful reading and judicious remarks.
\end{acknowledgements}


\begin{thebibliography}{}

  \bibitem[Boisse]{boisse} Boisse, I., Moutou, C., Vidal-Madjar, A. et al. 2009,
     \aap, 495, 959

  \bibitem[Bonfils et al. 2007]{bonfils} Bonfils, X., Mayor, M., Delfosse, X. et al. 2007,
      \aap, 474, 293
  
  \bibitem[]{} Bouchy, F., H\'ebrard, G., Udry, S. et al. 2009,
     \aap, 500, 853
  
  \bibitem[]{} Bruntt, H., Deleuil,M., Fridlund, M. et al. 2010, \aap, 519, 51
  
  \bibitem[2004]{desidera} Desidera, S., Gratton, R.G., Endl, M. et al. 2004,
    \aap, 420, L27-L30
  
  \bibitem[2007]{desort} Desort, M., Lagrange, A.-M., Galland, F. et al. 2007,
    \aap, 473, 983
  
  \bibitem[2009]{eggenberger} Eggenberger, A. \& Udry, S. 2010,
     EAS, 41, 27 
  
  \bibitem[2009]{forveille} Forveille, T., Bonfils, X., Delfosse, X. et al. 2009,
      \aap, 493, 645

  \bibitem[]{} Hatzes, A. P. 1999, ASPC, 185, 259
 
  \bibitem[]{} Hatzes, A. P. 2002, AN, 323, 392
      
  \bibitem[]{} Henry, G.W., Donahue, R.A. and Baliunas, S.L. 2002,
      \apj, 577, L111  

 \bibitem[1996]{howard} Howard, R.F. 1996, ARA\&A, 34, 75

  \bibitem[2008]{huelamo} Hu\'elamo, N., Figueira, P., Bonfils, X. et al. 2008,
      \aap, 489, L9-L13
      
   \bibitem[]{} Jenkins, J.S., Jones, H.R.A., Tinney, C.G. et al. 2006,
      \mnras, 372, 163   

  \bibitem[2000]{kurster} K\"urster, M., Endl, M., Els, S. and al., 2000,
      \aap, 353, L33-36
  
  \bibitem[2010]{lagrange} Lagrange, A.-M., Desort, M. and Meunier, N. 2010, \aap, 512, 38
       
     \bibitem[]{} Lanza, A.F., Bonomo, A.S., Moutou, C. et al, 2010, \aap, 520, 53
  
  \bibitem[]{} Lovis, C., S\'egransan, D., Mayor, M. et al. 2010, \aap, in press, arXiv:1011.4994
  
    \bibitem[Matthews et al.(2004)]{Matthews} Matthews, J.M. et al. (2004), Nature, 430, 51
  
   \bibitem[2003]{mayor} Mayor, M., Pepe, F., Queloz, D. and al., 2003,
      The Messenger, 114, 20M
   
   \bibitem[1924]{mc} McLaughlin D.B. 1924, \apj, 60, 22
   
   \bibitem[Melo et al. 2007] {melo} Melo, C., Santos, N.C., Gieren, W. et al. 2007, 
      \aap, 467, 721   
     
    \bibitem[2010]{meunier} Meunier, N., Desort, M. and Lagrange, A.-M. 2010, \aap, 519, 66
     
    \bibitem[]{} Paulson, D.B., Saar, S.H., Cochran, W.D. and Henry, G.W. 2004, 
         \aj, 127, 1644
   
    \bibitem[Queloz et al. 2001]{queloz}  Queloz, D., Henry, G. W., Sivan, J. P. et al. 2001,
       \aap, 379, 279 
       
   \bibitem[queloz]{queloz} Queloz, D., Bouchy, F., Moutou, C., Hatzes, A., H\'ebrard, G. et al.  2009, \aap, 506, 303
   
   \bibitem[]{}ÊRocha-Pinto, H.J. \& Maciel, W. J. 1998,
      \mnras, 298, 332
   
   \bibitem[1924]{rossiter} Rossiter R.A. 1924, \apj, 60, 15
   
   \bibitem[saar]{saar} Saar, S.H. \& Donahue, R. 1997,
       \apj, 485, 319
       
   \bibitem[]{} Saar, S.H., Huovelin, R.A., Osten, R.A. and Shcherbakov, A.G. 1997,
     \aap, 326, 741   
       
   \bibitem[]{} Saar, S. H. \& Osten R.A., 1997,
       MNRAS, 284, 803 
       
   \bibitem[]{} Santos, N.C., Mayor, M., Pepe, F. et al. 2000,
        \aap, 356, 599
        
   \bibitem[]{} Santos, N.C., Udry, S., Mayor, M. et al. 2003,
       \aap, 406, 373   
       
   \bibitem[]{} Santos, N. C., Gomes da Silva, J., Lovis, C. and Melo, C. 2010,
       \aap, 511, 45         
       
   \bibitem[setiawan]{setiawan} Setiawan, J., Henning, T., Launhardt, R., et al. 2008, Nature, 451, 38    
      
  \bibitem[]{} Valencia, D., Ikoma, M., Guillot, T. and Nettelmann, N. 2010, \aap, 516, 20     
      
  \bibitem[]{vauclair} Vauclair, S., Laymand, M., Bouchy, F. and al., 2008,
      \aap, 428L, 5V
      
  \bibitem[]{vogt} Vogt, S.S. \& Penrod, G.D. 1983,
         \pasp, 95, 565 
  

 \end{thebibliography}
\end{document}